\documentclass[12pt,a4]{iopart} 
\usepackage{colordvi,epsfig} 
\usepackage{pstricks}
\usepackage{psfrag}
\usepackage{pst-all}
\usepackage{subfigure}
\usepackage{amssymb}
\usepackage{bm}
\usepackage{cite}

\begin{document}

\title[Tearing mode with pressure flattening]{Tearing mode stability calculations with pressure flattening}
\author{C J Ham, J W Connor, S C Cowley, R J Hastie, T C Hender, Y Q Liu}
\address{EURATOM/CCFE Fusion Association, Culham Science Centre, Abingdon, Oxon, OX14 3DB, UK.}
\ead{christopher.ham@ccfe.ac.uk}
\date{\today}




\begin{abstract}
Calculations of tearing mode stability in tokamaks split conveniently into an external region, where marginally stable ideal MHD is applicable, and a resonant layer around the rational surface where sophisticated kinetic physics is needed.  These two regions are coupled by the stability parameter $\Delta'$.  Pressure and current perturbations localized around the rational surface alter the stability of tearing modes.  Equations governing the changes in the external solution and $\Delta'$ are derived for arbitrary perturbations in axisymmetric toroidal geometry.  The relationship of $\Delta' $ with and without pressure flattening is obtained analytically for four pressure flattening functions.  Resistive MHD codes do not contain the appropriate layer physics and therefore cannot predict stability directly.  They can, however, be used to calculate $\Delta'$.  Existing methods (Ham {\it et al.} 2012 {\it Plasma Phys. Control. Fusion} {\bf 54} 025009) for extracting $\Delta'$ from resistive codes are unsatisfactory when there is a finite pressure gradient at the rational surface and favourable average curvature because of the Glasser stabilizing effect.  To overcome this difficulty we introduce a specific pressure flattening function that allows the earlier approach to be used. The technique is first tested numerically in cylindrical geometry with an artificial favourable curvature. Its application to toroidal geometry is then demonstrated using the toroidal tokamak tearing mode stability code T7 (Fitzpatrick { \it et al.} 1993 {\it Nucl. Fusion} {\bf  33} 1533) which uses an approximate analytic equilibrium. The prospects for applying this approach to resistive MHD codes such as MARS-F (Liu {\it et al.} 2000 {\it Phys. Plasmas} {\bf 7} 3681) which utilize a fully toroidal equilibrium are discussed.
\end{abstract}
\submitto{\PPCF}
\maketitle

\section{Introduction} 
\label{sec:Intro}
Tearing modes in tokamaks can cause magnetic islands to form and these are expected to degrade the performance of burning plasma devices such as ITER or a future fusion power plant \cite{LaHaye06}. One therefore needs to identify operational regimes which minimise their impact. In cylindrical geometry the perturbed flux is assumed to have the form $\psi \sim e^{im\theta -in\phi}$ where $\theta$ and $\phi$ are the poloidal and toroidal angles respectively and $m$ and $n$ are poloidal and toroidal mode numbers. The stability of these modes for a given equilibrium can be determined from the dispersion relation obtained by matching solutions of the marginal ideal MHD equations away from a resonant surface, $m = n q(r_m)$ where $q$ is the safety factor, to those of an appropriate layer model in the vicinity of the resonant surface.  The information from the external ideal MHD equation in cylindrical geometry is characterised by a single quantity $\Delta'_m$ \cite{Furth63} and the layer solution is denoted by $\Delta_m(\omega)$ where $\omega$ is the frequency of the mode. The dispersion relation for the cylinder then becomes $\Delta'_m=\Delta_m(\omega)$. In toroidal geometry poloidal harmonics are coupled together so that the external MHD solution produces a relationship between $\Delta'_m$s. The external MHD equations therefore do not uniquely determine $\Delta'_m$ as they did in the cylinder. This relationship is given by Connor {\it et al.} \cite{Connor88}, although we use the notation of Fitzpatrick {\it et al.} \cite{Fitzpatrick93} where it is written as
\begin{equation}
|\bm{\Delta}-\bm{E}|=0
\end{equation}
with $\bm{\Delta}$ the matrix of layer responses and $\bm{E}$ the matrix corresponding to the external MHD equations. As an example, we can look at the system with two rational surfaces where
\begin{equation}
\bm{\Delta} = \left[ \begin{array}{cc} \Delta_m(\omega) & 0\\ 0 & \Delta_{m+1}(\omega) \end{array} \right].
\end{equation}
Alternatively, we can write this as \cite{Connor88} 
\begin{equation}\label{eq:IntroDisp}
(\Delta_{m}(\omega) -E_{11})(\Delta_{m+1}(\omega) -E_{22}) -E_{12}E_{21} =0 
\end{equation}
where $E_{11}$, $E_{22}$,  $E_{12}$ and $E_{21}$ give the information from external MHD and $\Delta_{m}(\omega)$ and $\Delta_{m+1}(\omega)$ are the two layer responses.

It has been shown by Cowley and Hastie \cite{Cowley88} that when diamagnetic terms are retained either $\Delta_{m}(\omega)$ or $\Delta_{m+1}(\omega)$ will be very large and so (\ref{eq:IntroDisp}) will reduce to two separate `cylinder like' cases
\begin{equation}
\Delta_{m}(\omega)= E_{11}, \qquad  \Delta_{m+1}(\omega) =E_{22},
\end{equation}
and a small sheared rotation will have the same effect. We can therefore define 
\begin{equation}
\Delta'_m=E_{11}, \quad \textrm{and} \quad \Delta'_{m+1}=E_{22}  
\end{equation}
by analogy with the cylinder. If there are $N$ resonant surfaces rather than just two there would then be $N$ of these `cylinder-like' dispersion relations which can be treated separately. Only if $\omega^*(r_m) +n\Omega(r_m)$ is very close to $\omega^*(r_{m+1})+n\Omega(r_{m+1})$ would we expect strong, linear, coupling of the two surfaces, where $\omega^*(r)$ is the electron diamagnetic drift frequency and $\Omega(r)$ is the toroidal angular velocity at $r$. For simplicity we drop the $m$ subscript from $\Delta'$ in the rest of the paper with the understanding that it is not a single quantity for a given equilibrium. 

Numerical codes for directly calculating $\Delta'$ in a torus have been developed, such as resistive PEST \cite{Pletzer91} and the toroidal stability code T7 \cite{Fitzpatrick93}, although these codes are not in routine use in the fusion community. However, the former, based on a finite element technique, has been applied to DIII-D \cite{Brennan02}. The second, T7, describes a shaped tokamak plasma cross section in terms of seven poloidal harmonics and solves the radial equations for the corresponding marginal ideal MHD harmonics as basis functions from which a toroidal $\Delta'$ is constructed. Since T7 is based on a large aspect ratio, weakly shaped analytic equilibrium its range of application is limited. 

Initial value codes such as FAR \cite{Charlton86, Hender87} or eigenvalue toroidal resistive MHD codes such as MARS-F \cite{Liu00}, in which the resonant layer model consists of the basic resistive MHD equations, can take full account of the complications of the toroidal geometry which couples the different poloidal harmonics. However, present and future large tokamaks require fully kinetic descriptions of the linear, or non-linear, physics in the resonant layer. Our objective is to extract the information on $\Delta'$ from such resistive MHD codes so that it can be used together with realistic layer models for a proper determination of tearing mode stability.

In the absence of a pressure gradient at the resonant surface two approaches to solving this problem have been presented in Ham {\it et al.}\cite{Ham12b}. Firstly one can deduce the $\Delta'$ from the tearing mode growth rate, $\gamma$, calculated by a resistive MHD code using the known dispersion relation for the resistive MHD model. Alternatively, one can use a resistive MHD code to obtain a set of basis functions from which $\Delta'$ can be constructed. This involves a set of fully reconnected solutions (i.e. continuous at the relevant resonant surfaces and thus  containing the large solution in the sense of Newcomb \cite{Newcomb60}) and a set of small solutions, again in the sense of Newcomb, emerging from the various resonant surfaces which can be combined to satisfy the appropriate boundary conditions at the magnetic axis and plasma edge and used to deduce $\Delta'$.

However both approaches have difficulties when there is a finite pressure gradient and favourable average curvature at a resonant surface due to the `Glasser effect' \cite{Glasser75}. The internal layer solution to be matched to the outer MHD solution for a large aspect ratio circular plasma, with pressure gradient at the rational surface, is
\begin{equation}\label{eq:IntroGGJ}
\Delta_{GGJ}(\gamma)=2.12 A (\gamma \tau_A)^{5/4} \left[1-\frac{\pi}{4} D_R B (\gamma \tau_A)^{-3/2}  \right],
\end{equation}
\begin{eqnarray}
A &\equiv& \left(\frac{nq'}{q} \right)^{-1/2}(1+2q^2)^{1/4}\left( \frac{\tau_R}{\tau_A}\right)^{3/4} \\
B &\equiv& \left(\frac{nq'}{q} \right)(1+2q^2)^{-1/2}\left( \frac{\tau_R}{\tau_A}\right)^{-1/2}
\end{eqnarray}
where $\gamma$ is the tearing mode growth rate which could be a complex number, a prime represents a derivative with respect to the minor radius, and $D_R$ is the resistive interchange stability index which depends on field geometry only (not on $\tau_R/\tau_A$). The term involving $D_R$ comes from the favourable average curvature and it is proportional to the pressure gradient. The value of $D_R$ is normally a small negative number. It is this term that produces a stabilizing effect so that the tearing mode is stable below some critical value $\Delta'=\Delta'_c>0$. The effect is stronger when the resistivity is low, in fact $\Delta'_c \sim (\tau_R/\tau_A)^{1/3}$. 

In this situation with asymptotically low values of plasma resistivity, an initial value resistive MHD code predicts stability even if $\Delta'>0$, the simple low pressure cylindrical instability criterion. Thus one cannot deduce the $\Delta'$ from $\gamma$ using an initial value resistive MHD code such as FAR. In principle an eigenvalue code like MARS-F can obtain stable eigenvalues and values for $\Delta'$ deduced by solving the relevant resistive layer equations. Indeed for certain classes of equilibria where an analytic dispersion relation results \cite{Glasser75}, this has been demonstrated by Liu {\it et al.} \cite{Liu12}, although it is necessary to follow the eigenvalue into the stable region with some care. Furthermore, for general equilibria a numerical solution of the equations describing the resistive layer becomes necessary, adding additional complication. The second method involves calculating the solution at a resonant surface with $\gamma=0$, but in the presence of a finite pressure gradient this solution, as shown in (\ref{eq:IntroGGJ}), gives $\Delta' \to \infty$ (i.e. the resonant surface is completely screened from an external perturbation \cite{Glasser75, Coppi66}, see Appendix A) and one cannot determine the large solution, so this technique also fails.

We can overcome this problem by introducing a pressure flattening function at the rational surfaces which will remove the Glasser effect. In earlier work, Bishop {\it et al.} \cite{Bishop91} introduced a localised pressure perturbation at the resonant surface in order to assess the sensitivity of $\Delta'$ to such effects. Considering a cylindrical model, but with an artificial favourable average curvature, and a specific form for the pressure perturbation that flattened the pressure profile at the resonant surface, these authors calculated analytically the relationship between $\Delta'_{\infty}$, the value of $\Delta'$ corresponding to the finite pressure gradient, and $\Delta'_{0}$, the value with flattened pressure. This procedure indicates how one might overcome the Glasser effect and deduce $\Delta'$ in toroidal geometry with a finite pressure gradient using a toroidal resistive MHD code. Whereas the calculation of Bishop {\it et al} \cite{Bishop91} was motivated by the need to understand the sensitivity of $\Delta'$ to possible, but realistic, details of the behaviour of the pressure gradient near the rational surface, we are free to design this behaviour so as to facilitate the determination of $\Delta'_{\infty}$ in terms of $\Delta'_0$ without regard to its physical accessibility.

It should be noted that in codes like resistive PEST and T7, as the pressure gradient at the resonant surface increases, the two Mercier indices move apart and it becomes difficult to extract the large and small solutions, so methods for flattening the pressure also have a role to play in extending the applicability of such a tearing mode stability code.

In Section 2 we derive the appropriate second order differential equation that governs the tearing mode solution in the vicinity of the resonant surface and in the presence of the pressure flattening perturbation for full toroidal geometry. Some details of this calculation are shown in Appendix B. The analysis also includes a similar perturbation to the $q$-profile and a specific calculation of the effect of this on $\Delta'$ is presented in Appendix C. Our choice of pressure flattening perturbation is presented in Section 3 and the relationship between the corresponding two values of $\Delta'$ is calculated analytically for this case. In Appendix D we present the corresponding results for a generalization of the Bishop {\it et al.} \cite{Bishop91}  model and another variant motivated by similar ideas to the one discussed in Section 2.

Examples of the implementation of these models in cylindrical geometry, with an artificial favourable average curvature, are described in Section 4, while Section 5 demonstrates the application to toroidal geometry using the toroidal tearing mode stability code T7. In the rest of this paper we focus on an equilibrium with a single rational surface. It will be trivial to generalize to an equilibrium with many rational surfaces. In contrast to the cylindrical case where the pressure gradient appears only at one point in the Newcomb equation, in the torus it affects various toroidal couplings and metric coefficients, for instance through the Shafranov shift. This suggests two options: (i) to self consistently change the equilibrium throughout or; (ii) to act as a `Maxwell Demon' and intervene only in the equation governing the resonant harmonic. Both options are investigated and compared. In the conclusion we discuss the prospects for applying this approach to fully toroidal resistive MHD codes like MARS-F \cite{Liu00} to obtain accurate values of $\Delta'$ in the presence of a pressure gradient.

\section{The Localised Tearing Mode Equation with Pressure Flattening}

In this section we investigate how a localized equilibrium perturbation affects the stability of tearing modes. We shall consider the situation where this perturbation extends over a region of width $\delta$ ($\delta \ll r_a$, the plasma minor radius) which includes the rational surface $q=m/n$. The non-ideal MHD layer where resistivity, diamagnetic drifts and other kinetic effects are important is of width $w$, and we shall assume that $w\ll \delta$. We can then treat the perturbed region as an ideal MHD boundary layer: the non-ideal layer is a boundary layer nested within this ideal boundary layer. Our object is to connect solutions $|r-r_m| \gg \delta$ to those for $w \ll |r-r_m| \ll \delta$. We introduce $t=(r-r_m)/\delta$ where we are interested in the regimes $t\gg1$ and $t\ll 1$. 

\begin{figure}
\centering
\psfrag{A}{$t=(r-r_m)/\delta$}
\psfrag{B}{$\hat{p}(t)$}
\psfrag{C}[Bl][Bc][1][90]{Outer region}
\psfrag{D}[Bl][Bc][1][90]{Transition region}
\psfrag{E}[Bl][Bc][1][90]{Flattened region}
\includegraphics[width=0.8\textwidth]{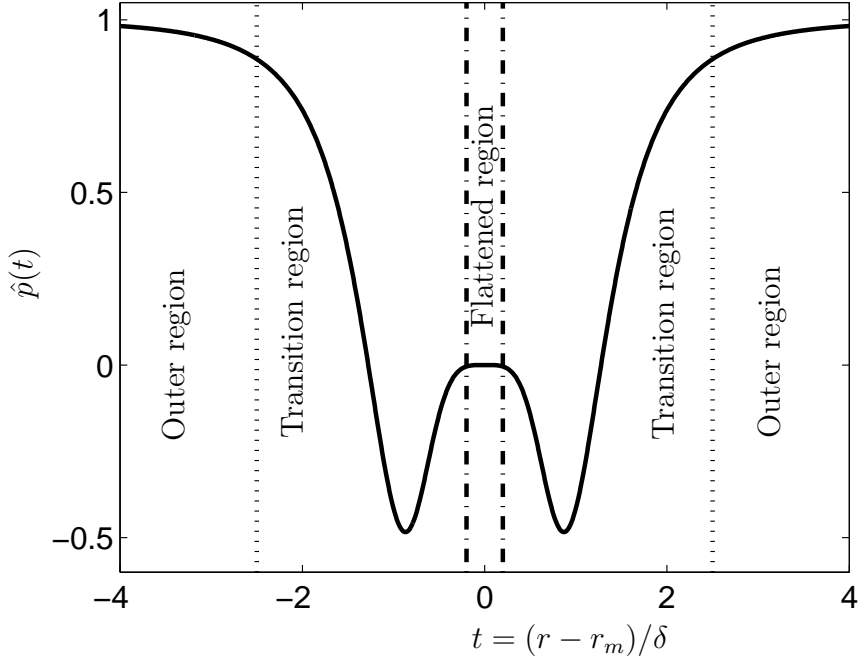}
\caption{The pressure flattening form function, which will be used later, showing the regions of interest. The dashed lines at $t=\pm 2.5$ represent the transition of the regions from the outer region, $|t|>2.5$, to the transitional region, $0.2<|t|<2.5$. The dash-dot lines at $t=\pm 0.2$ show the transition to the region where the pressure gradient has been flattened to zero.}
\label{fig:SCCplot}
\end{figure}

Figure \ref{fig:SCCplot} shows an example of a pressure gradient form factor $\hat{p}(t)$ such that if $p'^{(0)}(r)$ is the unperturbed pressure gradient then $p'^{(0)}(r)\hat{p}(t)$ is the pressure gradient with perturbation. This is the pressure flattening function that is used in Section 3. The three regions of interest can be seen in this plot: a region far from the rational surface where the pressure is not changed; a transitional region; and a region where the pressure has been completely flattened just around the rational surface. We seek to connect the outer region with the flattened region. 

We now derive a second order differential equation for $\psi_m$ in the region $r_a \gg |r-r_m| \gg w$. This equation connects the asymptotic solutions for $\delta \gg |r-r_m|$, to those for $ |r-r_m| \gg \delta$. We expand the marginal MHD equations around the surface $q=m/n$:
\begin{eqnarray}
\label{eq:Exp1}
\hspace*{-1cm}\frac{\partial}{\partial r}\left[ \left(\frac{\partial}{\partial \theta} -inq\right)y \right] &=& \frac{\partial}{\partial \theta}
\left(Q\frac{\partial Z}{\partial \theta} \right) + SZ - \frac{\partial}{\partial \theta}\left[T\left(\frac{\partial}{\partial \theta} -inq\right)y +Uy \right]\\
\label{eq:Exp2}\left(\frac{\partial}{\partial \theta} -inq\right) \frac{\partial Z}{\partial r} &=& U \frac{\partial Z}{\partial \theta} - \left(\frac{\partial}{\partial \theta} -inq\right)T^* \frac{\partial Z}{\partial \theta}+ Xy \nonumber \\
&+& W \left(\frac{\partial}{\partial \theta} -inq\right)y - \left(\frac{\partial}{\partial \theta} -inq\right)V\left(\frac{\partial}{\partial \theta} -inq\right)y
\end{eqnarray}
where the dependent variables are $y=R_0f\bm{\xi}.\nabla r$ and $Z=R^2\delta \bm{B}.\nabla\phi/B_0$, with $\bm{\xi}$ the plasma displacement and $\delta {\bf B}$ the perturbed magnetic field, while $Q,S,T,U,V,W,$ and $X$ are equilibrium quantities (with $T^*$ the complex conjugate of $T$) defined in Connor {\it et al.} \cite{Connor88}:
\begin{eqnarray}
Q&=& R_0/\left(inr|\nabla r|^2  \right); \qquad S=inr/R_0 \nonumber \\
U&=& \frac{p'}{B^2_0f^2}\frac{R^2}{R^2_0}\frac{1}{|\nabla r|^2}; \qquad V= \frac{R_0}{r|\nabla r|^2}\left[\frac{in}{R^2}
+\frac{1}{in}\left(\frac{g'}{f}  \right)^2 \right] \nonumber \\
T&=& \frac{1}{|\nabla r|^2}\left[\nabla\theta\cdot \nabla r - \frac{R_0}{inr}\frac{g'}{f}  \right] \\
W&=& 2\frac{g'}{g}U - \frac{d}{dr}\left(\frac{g'}{f} \right) \nonumber \\
X&=& \frac{inp'}{B^2_0 f^2}\frac{r}{R}\left[ \frac{\partial}{\partial \theta} \left(T^*\frac{R^2}{R^2_0} \right)+
\frac{\partial}{\partial r}\left(\frac{R^2}{R^2_0} \right)-\frac{R^2}{R^2_0}\frac{r}{f}\frac{d}{dr}\left(\frac{f}{r} \right)
-U\frac{R^2}{R^2_0}  \right]. \nonumber
\end{eqnarray}
The magnetic field has been represented as
\begin{equation}
\label{eq:Bs2}
{\bf B} = R_0B_0(f(r) \nabla \phi \times \nabla r + g(r) \nabla \phi),
\end{equation}
where $\phi$ is the toroidal angle, $p$ is the pressure, a prime denotes the derivative with respect to $r$, $R_0$ is the major radius at the magnetic axis and $B_0$ is the magnetic field there, following Connor {\it et al.} \cite{Connor88}

To facilitate the expansion we represent $q(r)\equiv \frac{m}{n}+\hat{\delta}\Delta q(t)$ where $t=(r-r_m)/\delta$ and $\hat{\delta}=\delta/r_m$ and expand in $\hat{\delta} \ll 1$. ($\Delta q(t)$ is related to the perturbation $q^{(1)}(t)$ appearing in (\ref{eq:qAppB}) by $\Delta q(t)=r_mtq^{(0)'}(r_m)+q^{(1)}(t)$.) 
Furthermore we write 
\begin{eqnarray}
p&=&p^{(0)}(r)+\hat{\delta} p^{(1)}(t)\\
f&=&f^{(0)}(r)+\hat{\delta} f^{(1)}(t)\\
g&=&g^{(0)}(r)+\hat{\delta} g^{(1)}(t)
\end{eqnarray}
and
\begin{equation}
y=y^{(0)}+\hat{\delta} y^{(1)} +\hat{\delta}^2y^{(2)}+..., \quad Z=Z^{(0)} +\hat{\delta} Z^{(1)}+ \hat{\delta}^2 Z^{(2)}+...
\end{equation}

In leading order $(O(\hat{\delta}^{-1}))$ (\ref{eq:Exp1}) and (\ref{eq:Exp2}) yield
\begin{eqnarray} 
\frac{\partial}{\partial t} \left( \frac{\partial}{\partial \theta}-im \right)y^{(0)}&=&0\\
\left( \frac{\partial}{\partial \theta}-im \right)\frac{\partial Z^{(0)}}{\partial t}&=&-\frac{1}{f^{(0)}}\frac{d^2 g^{(1)}(t)}{dt^2} \left( \frac{\partial}{\partial \theta}-im \right)y^{(0)}
\end{eqnarray}
since 
\begin{equation}
W_{-1}=\frac{1}{rf}\frac{d^2 g^{(1)}}{dt^2} \nonumber
\end{equation}
and where $r$ is evaluated at $r_m$.
These equations have solutions of the form
\begin{eqnarray}
y^{(0)}&=& \bar{y}^{(0)}(t)e^{im\theta}+ \bar{\bar{y}}^{(0)}(r,\theta)\\
Z^{(0)}&=& \bar{Z}^{(0)}(t)e^{im\theta}+ \bar{\bar{Z}}^{(0)}(r,\theta)- \frac{1}{f^{(0)}}\frac{d g^{(1)}}{dt}\bar{\bar{y}}^{(0)}(r,\theta)
\end{eqnarray}
where $\bar{y}^{(0)}$ and $\bar{Z}^{(0)}$ contain the large and small solutions as $\Delta q \to 0$, and $\bar{\bar{y}}^{(0)}$ and $\bar{\bar{Z}}^{(0)}$ represent the regular part of the solution which is present in toroidal calculations of this nature.

Proceeding to next order in $\hat{\delta}$, i.e. $O(\hat{\delta}^0)$, the follow equations emerge:

\begin{eqnarray}
\hspace*{-1cm}&&\frac{1}{r}\frac{\partial}{\partial t}\left[\left(\frac{\partial}{\partial \theta}-im \right)y^{(1)}  \right]
-\frac{in}{r} \frac{\partial}{\partial t}(\Delta qy^{(0)})+\frac{\partial}{\partial r}\left[\left(\frac{\partial}{\partial \theta}-im \right)\bar{\bar{y^{(0)}}}  \right]\nonumber \\
&&=\frac{\partial}{\partial \theta}\left[Q \frac{\partial Z^{(0)}}{\partial \theta}  \right] + SZ^{(0)}- \frac{\partial}{\partial \theta} \left[T \left(\frac{\partial}{\partial \theta}-im \right)\bar{\bar{y^{(0)}}} +U y^{(0)}  \right]
\end{eqnarray}
and
\begin{eqnarray}
&&\hspace*{-2cm}\left(\frac{\partial}{\partial \theta}-im \right)\frac{1}{r}\frac{\partial Z^{(1)}}{\partial t} -\frac{in}{r}\Delta q \frac{\partial Z^{(1)}}{\partial t} = U \frac{\partial Z^{(0)}}{\partial \theta} - \left(\frac{\partial}{\partial \theta}-im \right)T^*\frac{\partial Z^{(0)}}{\partial \theta} + Xy^{(0)}  \\
&&\hspace*{-2cm}+ W_{-1} \left[\left(\frac{\partial}{\partial \theta}-im \right)y^{(1)} -in\Delta qy^{(0)}  \right] +W_0 \left(\frac{\partial}{\partial \theta}-im \right) y^{(0)}-\left(\frac{\partial}{\partial \theta}-im \right)V \left(\frac{\partial}{\partial \theta}-im \right)y^{(0)} \nonumber
\end{eqnarray} 

To obtain equations for $y^{(0)}$ and $Z^{(0)}$, we annihilate $y^{(1)}$ and $Z^{(1)}$ in these equations by operating with $\frac{1}{2\pi}\oint d\theta e^{-im\theta}$. The resulting equations take the following form
\begin{equation}
\label{eq:qy0}
\frac{1}{r}\frac{d}{dt}(\Delta q \bar{y}^{(0)})= -\frac{r}{R_0 f^2} \left \langle \frac{B^2}{B^2_0}\frac{R^2}{R^2_0}\frac{1}{|\nabla r|^2} \right\rangle
\bar{Z}^{(0)} +\frac{m}{n}\left\langle \frac{R^2}{R^2_0}\frac{1}{|\nabla r|^2} \right\rangle \beta'_p\bar{y}^{(0)}
\end{equation}
and 
\begin{eqnarray}
&&\frac{\Delta q}{r} \frac{d\bar{Z}^{(0)}}{dt} = -\frac{g^{(1)''}}{f}\Delta q \bar{y}^{(0)}-\beta'_p\frac{m}{n}\left \langle \frac{R^2}{R^2_0}\frac{1}{|\nabla r|^2} \right \rangle \bar{Z}^{(0)} \nonumber \\
&&- \beta'_p \frac{r}{R_0}\left \langle \frac{\partial}{\partial r}\left( \frac{R^2}{R^2_0} \right) - \frac{R^2}{R^2_0}\frac{r}{f} \frac{d}{dr} \left( \frac{f}{r}\right) - \beta'_p \frac{R^4}{R^4_0} \frac{1}{|\nabla r|^2} \right \rangle \bar{y}^{(0)}\label{eq:DqZ0}
\end{eqnarray}
where 
\begin{equation}
\beta'_p=\frac{p'}{B^2_0f^2}\equiv \frac{1}{B^2_0f^2}\left[ \frac{dp^{(0)}}{dr}+ \frac{dp^{(1)}(t)}{dt}\right] \equiv \beta^{(0)'}_0 +\beta'^{(1)}_p, \quad \frac{df}{dr}= \frac{df^{(0)}}{dr}+ \frac{df^{(1)}}{dt} 
\end{equation}
and we have suppressed the inhomogeneous terms involving $\bar{\bar{y}}^{(0)}$ and $\bar{\bar{Z}}^{(0)}$, since these merely determine the $m$th harmonic of the continuous solution in $\bar{y}_0$ and $\bar{Z}_0$ and do not influence the behaviour of the large and small solutions in the region adjacent to the rational surface.

To proceed further we use equilibrium relations (\ref{eq:fAppB}), (\ref{eq:oqAppB}) and (\ref{eq:AppBFinal}) to simplify the last term in (\ref{eq:DqZ0}). Thus
\begin{eqnarray}
&& \left \langle \frac{\partial}{\partial r} \left(\frac{R^2}{R^2_0} \right) - \frac{R^2}{R^2_0}\frac{r}{f}\frac{d}{dr} \left(\frac{f}{r} \right) - \beta'_p \frac{R^4}{R^4_0}\frac{1}{|\nabla r|^2} \right \rangle \\
&=& \frac{d}{dr}\left \langle \frac{R^{(0)2}}{R^2_0} \right \rangle - \left \langle \frac{R^{(0)2}}{R^2_0} \right \rangle \frac{r}{f^{(0)}} \frac{d}{dr} \left(\frac{f^{(0)}}{r} \right)- \left \langle \frac{R^4}{R^4_0}\frac{1}{|\nabla r|^2} \right \rangle \frac{1}{B^2_0 f^{(0)2}}\frac{dp^{(0)}}{dr} \nonumber \\
&+& \frac{g^{(0)2}}{r} \left \langle \frac{R^{(0)2}}{R^2_0|\nabla r|^2} \right \rangle \left \langle \frac{R^{(0)2}B^2}{R^2_0B^2_0 |\nabla r|^2} \right \rangle^{-1}\left[\frac{1}{q^{(0)}} \frac{d q^{(1)}}{dt}-\frac{1}{B^2_0 f^{(0)2}} \left \langle \frac{R^{(0)2}}{R_0^2 |\nabla r|^2} \right \rangle \frac{d p^{(1)}}{dt} \right] \nonumber
\end{eqnarray}
Finally, defining $\psi_m=\Delta q(t) \bar{y}^{(0)}(t)$, and eliminating $\bar{Z}^{(0)}$ from (\ref{eq:qy0}) and (\ref{eq:DqZ0}), we obtain a simple second order ODE for $\psi(t)$ in the form
\begin{equation}
\label{eq:q2final}
(\Delta q)^2 \frac{d^2\psi_m}{dt^2}-\psi_m\left [\Delta q \frac{d^2 (\Delta q)}{dt^2}+\left(\beta^{(0)'}_p +\beta'^{(1)}_p \right)K \right ]=0
\end{equation}
where $K$ is independent of $t$ and is given by
\begin{eqnarray}
\hspace*{-1.5cm}K&=& \left(\frac{r^2}{R_0}\right)^2\frac{1}{f^{(0)2}}\left \langle \frac{B^2 R^{(0)2}}{B_0^2R_0^2|\nabla r|^2} \right \rangle\nonumber \\
\hspace*{-1.5cm}&\times&\left \{\frac{d}{dr} \left \langle \frac{R^{(0)2}}{R_0^2}\right \rangle - \left\langle \frac{R^{(0)2}}{R_0^2} \right\rangle \frac{r}{f^{(0)}} \frac{d}{dr} \left(\frac{f^{(0)}}{r}\right) -\left\langle \frac{R^{(0)4}}{R_0^4 |\nabla r|^2} \right\rangle \frac{1}{B_0^2 f^{(0)2}}\frac{dp^{(0)}}{dr} \right \}  \\
\hspace*{-1.5cm}&+& r^2q^{(0)2}\beta^{(0)'}_p \left \langle \frac{R^2}{R_0^2|\nabla r|^2} \right\rangle ^2 - r^2q^{(0)}q^{(0)'}\left \langle \frac{R^{(0)2}}{R_0^2} \frac{1}{|\nabla r|^2} \right \rangle \nonumber
\end{eqnarray}
or, in standard notation,
\begin{equation}
K\equiv K^{(0)}= D_{M \infty}  \frac{(rq^{(0)'})^2}{\beta^{(0)'}_p}
\end{equation}
where the Mercier stability criterion \cite{Mercier60} (for the unperturbed equilibrium) takes the form $\frac{1}{4} +D_{M \infty}>0$. We will assume that this equilibrium is Mercier stable.

We now look at (\ref{eq:q2final}) and find the solutions in the limits $\delta \gg |r-r_m|$ and $ |r-r_m| \gg \delta$. In the limit $t\gg1$, which is equivalently $\delta \ll |r-r_m|$, (\ref{eq:q2final}) becomes
\begin{equation}
\label{eq:q2tgg1}
t^2 \frac{d^2 \psi_m}{dt^2}-\psi_m D_{M\infty}=0.
\end{equation}
The solution of this equation is
\begin{eqnarray}
\label{eq:SolInf}
\psi^{+,-}_{m\infty} &=& a^{+,-}_{\infty}|t|^{\nu_{L}} + b^{+,-}_{\infty}|t|^{\nu_{S}}\nonumber \\
&=& \alpha^{+,-}_{\infty} |x|^{\nu_{L}} + \beta^{+,-}_{\infty} |x|^{\nu_{S}},
\end{eqnarray} 
where we rescale to the variable $x=(r-r_m)/r_m$, $\psi_{m \infty}$ is the component of the helical flux function that is resonant at $r_m$,  
\begin{equation}
\nu_L=\frac{1}{2}(1-\sqrt{(1+4D_{ M \infty})}) \quad \textrm{and} \quad \nu_S=\frac{1}{2}(1+\sqrt{(1+4D_{ M \infty})}) 
\end{equation}
are the large and small Mercier indices respectively and $\alpha^{+,-}_{\infty}$ and $\beta^{+,-}_{\infty}$ are the coefficients of the corresponding large and small solutions to the right $(+)$ or the left $(-)$ of $r_m$.

In the limit $t\ll1$, which is equivalently $\delta \gg |r-r_m|$, (\ref{eq:q2final}) becomes
\begin{equation}
\label{eq:q2tll1}
\frac{d^2 \psi_m}{dt^2}=0.
\end{equation}
The solution to this equation is
\begin{eqnarray}
\label{eq:Sol0}
\psi^{+,-}_{m0} &=& a^{+,-}_{0} + b^{+,-}_{0}|t|\nonumber \\
&=& \alpha^{+,-}_0 + \beta^{+,-}_0 |x|.
\end{eqnarray}

Since this is a linear problem we can write matrix equations relating $\alpha^{+,-}_{0}$ to $\alpha^{+,-}_{\infty}$ and $\beta^{+,-}_{0}$ to $\beta^{+,-}_{\infty}$ as
\begin{equation}
\left( \begin{array}{c} \alpha^+_0 \\ \beta^+_0  \end{array} \right) = \bm{M}^+ \left( \begin{array}{c} \alpha^+_{\infty} \\ \beta^+_{\infty} \end{array} \right) 
\end{equation}
and 
\begin{equation}
\left( \begin{array}{c} \alpha^-_0 \\ \beta^-_0 \end{array} \right) = \bm{M}^- \left( \begin{array}{c} \alpha^-_{\infty} \\ \beta^-_{\infty} \end{array} \right). 
\end{equation}
The object here is to find $\bm{M}^+$ and $\bm{M}^-$. We take $p'(t)$ to be symmetric about $t=0$ so that $\bm{M}^+ =\bm{M}^-$. The elements of $\bm{M}^{+,-}$ can be investigated by considering two basis functions at large $t$: the pure large solution $\psi_L(t)\sim |t|^{\nu_L}+...$; and the pure small solution $\psi_S(t)\sim|t|^{\nu_S}+...$. These basis functions in general generate large and small solutions at small $t$
\begin{eqnarray}
\psi_S(t) &\to& c_S + d_S |t|, \\
\psi_L(t) &\to& c_L + d_L |t|, 
\end{eqnarray}
where $c_L, c_S, d_L$ and $d_S$ are determined by the flattening function. The matrix $\bm{M}^{+,-}$ can then be written as
\begin{equation}\label{eq:ExplicitM}
\bm{M}^{+,-}=\left( \begin{array}{cc} c_L \hat{\delta}^{\nu_L} & c_S \hat{\delta}^{\nu_S}  \\
 d_L \hat{\delta}^{\nu_L-1} & d_S \hat{\delta}^{\nu_S-1} \end{array} \right).
\end{equation} 
We may define $\Delta'_0$ from the solutions to (\ref{eq:q2final}) given above in the region $t \gg 1$ as
\begin{equation}
\Delta'_0= \frac{\beta^{+}_{0}}{\alpha^{+}_{0}}+\frac{\beta^{-}_{0}}{\alpha^-_{0}}.
\end{equation}

We may also define $\Delta'_{\infty}$ from the solutions to (\ref{eq:q2final}) in the region $t \gg 1$ as
\begin{equation}
\Delta'_{\infty}=\frac{\beta^{+}_{\infty}}{\alpha^{+}_{\infty}}+\frac{\beta^{-}_{\infty}}{\alpha^-_{\infty}}
\end{equation}
where the superscripts $+(-)$ again indicate $r>r_m$ $(r<r_m)$ respectively. 
We can link $\Delta'_0$ to $\Delta'_{\infty}$ by using (\ref{eq:ExplicitM}) 
\begin{equation}
\frac{\beta_0}{\alpha_0}=\frac{d_L \hat{\delta}^{-1} + d_S\left(\frac{\beta_{\infty}}{\alpha_{\infty}}\right ) \hat{\delta}^{-2\nu_L}} {c_L + c_S\left(\frac{\beta_{\infty}}{\alpha_{\infty}}\right ) \hat{\delta}^{\nu_S-\nu_L}}.
\end{equation}
If we take $\hat{\delta}$ sufficiently small that $c_L \gg c_S\frac{\beta_{\infty}} {\alpha_{\infty}} \hat{\delta}^{\nu_S-\nu_L}$ then we can write a general expression for $\Delta'_{0}$ in terms of $\Delta'_{\infty}$
\begin{equation}\label{eq:Delta0Gen}
\Delta'_0=u \hat{\delta}^{-2\nu_L}\Delta'_{\infty} +v/\hat{\delta}
\end{equation}
where $u= (d_Sc_L -d_Lc_S)/c_L^2$ and $ v=d_L/c_L$ are constants depending on the perturbed quantities. Note that for $\hat{\delta} \ll 1$ and $v\neq 0$ the second term on the right hand side of (\ref{eq:Delta0Gen}) dominates.

Our purpose is to use analytic means to determine $\Delta'_{\infty}$ from a numerical calculation of $\Delta'_0$. In order to validate our method we will consider situations in which the solution for $t\gg 1$ can be calculated directly: numerically, as in cylindrical geometry or using the T7 code in a large aspect ratio torus. 

The main result of this section is contained in (\ref{eq:q2final}) which describes how the large and small singular solutions behave in the vicinity of a rational surface under the influence of arbitrary localized perturbations to the safety factor and the pressure. A remarkable feature of (\ref{eq:q2final}) is that, although the magnetic well, or average curvature, may be modified by local currents, the Mercier index ($\beta'_p K$) only responds to pressure perturbations in $\beta'^{(1)}_p$; therefore all perturbation effects cancel in $K$.
 
\section{The Pressure Flattening Function and the Effect on $\Delta'$} 

The effect on the equation for $\psi_m$, i.e. (\ref{eq:q2final}), of perturbing the pressure and current in a general torus has been calculated in the previous section. In this section we use this equation with a perturbation that flattens the pressure at the resonant surface to allow us to calculate the $\Delta'$ for an equilibrium with pressure. Considering the situation in which there is no change in the $q$ profile, i.e. $q^{(1)}(t)=0$, (\ref{eq:q2final}) takes the form 
\begin{equation}
\label{eq:Start}
t^2 \frac{d^2\psi_m}{dt^2}=D(t)\psi_m
\end{equation}
where $D(t) = D_{M\infty} p'(t)/p'_0$. Bishop {\it et al.} \cite{Bishop91} used a particular  function to flatten the pressure at the rational surface.
\begin{equation}
\label{eq:BisPres}
p'(t)=p'_0\frac{t^2}{t^2+1}
\end{equation}   
where $r_m$ is the location of the rational surface, $t=(r-r_m)/\delta$ is the local radial coordinate and, $\hat{\delta}=\delta/r_m \ll 1$. Solving (\ref{eq:Start}) produced an analytic relationship for $u$ and $v$ in (\ref{eq:Delta0Gen})
\begin{equation}
\Delta'_0=\hat{\delta}^{-2\nu_L}\Delta'_{\infty}-\frac{\pi D_{M\infty}}{2\hat{\delta}},
\end{equation}
which is valid for low $\beta$ equilibria where $D_{M\infty} \ll 1$. Typically $D_{M\infty} >0$ for tokamaks with $q>1$. For $\hat{\delta} \ll 1$, $D_{M\infty} >0$ and $\Delta'_{\infty}$ finite this yields a large negative $\Delta'_0$ and stable tearing modes. Indeed $\Delta'_0$ is insensitive to $\Delta'_{\infty}$ except where $\Delta'_{\infty}$ is very large.  

In the present work we produce a new flattening function which leads to $v=0$ by carefully designing the behaviour of the large Mercier solution as it approaches the relevant mode rational surface and deducing the corresponding form of $p'(t)$. The large solution is taken to be exactly
\begin{equation}
\label{eq:psiL}   
\psi_L(t)= (1+|t|^{2l})^{\frac{\nu_L}{2l} }
\end{equation}
where the index $l$ can be chosen to produce sharper, or gentler, variations with $t$. The resulting $D(t)$ is calculated by requiring solution (\ref{eq:psiL}) to satisfy (\ref{eq:Start}), taking $q^{(1)}(t) =0$:
\begin{equation}
D^{(1,l)}(t)=\frac{t^{2l}}{(1+t^{2l})^2}  \left [ t^{2l} D_{M \infty}+\nu_L (2l-1)\right ]
\end{equation}

Clearly, (\ref{eq:psiL}) has the property that a large Mercier solution approaching the mode rational surface simply changes its functional form (from $|t|^{\nu_L}$  to  $|t|^0$) without changing its amplitude as it passes through the pressure flattened region, $|t|\sim O(1)$. There is no componant of the small solution in $\psi_L$ as $t\to 0$ i.e. $d_L=0$. The behaviour of the small solution is more complex. However, knowing one solution, the second solution of (\ref{eq:Start}), can be constructed as
\begin{equation}
\label{eq:psiS}
\hspace*{-1.5cm}\psi_S(t)= (1+|t|^{2l})^{\frac{\nu_L}{2l}}  |t|^{1-2 \nu_L}-(1-2\nu_L)(1+|t|^{2l})^{\frac{\nu_L}{2l}} \int_{|t|} ^\infty dx \left [ (1+x^{2l})^{\frac{-\nu_L}{l}}-x^{-2\nu_L} \right ].
\end{equation}
In constructing this $\psi_S(t)$ we have chosen a constant of integration in such a way that, as $|t|\to \infty$, $\psi_S$ is purely small solution. As this solution passes through the pressure flattened region it emerges, at small $t$, as a mix of large and small Mercier solutions.
Specifically, for $|t|\ll1$,
\begin{equation}
\psi_S \sim   (2 \nu_L -1) (I_0 -|t|)
\end{equation}
where $I_0(l,\nu_L)$ is the infinite integral:
\begin{equation}
I_0 = \int_0^\infty dx\left [ (1+x^{2l})^{\frac {-\nu_L}{l}} - x^{-2\nu_L} \right ] = \frac{\Gamma(1+\frac{1}{2l}) \Gamma(\frac{2\nu_L -1}{2l})}{\Gamma(\frac{\nu_L}{l})}.
\end{equation}
We use these results to show that $c_L=1$, $d_L=0$, $c_S=(2 \nu_L-1)I_0$ and $d_S=(1-2\nu_L)$. This leads to the following expression for the quantity $\Delta'_{\infty}$:
\begin{equation}
\label{eq:Dapprox}
\Delta'_{\infty} = \frac{\hat{\delta}^{2\nu_L}}{1-2\nu_L} \Delta'_0.
\end{equation}
This relationship has the general form (\ref{eq:Delta0Gen}) but importantly $v=0$ so there is no offset and $\Delta'_0$ is directly proportional to $\Delta'_{\infty}$. This direct relationship means that if $\Delta'_0$ is positive so is $\Delta'_{\infty}$. We can use (\ref{eq:IntroGGJ}) to show that with pressure flattening
\begin{equation}
\gamma \tau_A = 0.548 \Delta_0'^{4/5} \left(\frac{nq'}{q} \right)^{2/5} \left( \frac{\tau_A}{\tau_R} \right)^{3/5} (1+2q^2)^{-1/5} 
\end{equation}
and so we can use a resistive code when $\Delta'_0>0$ to find $\Delta'_{\infty}$. In Appendix D we provide the links between $\Delta'_{\infty}$ and $\Delta'_0$ for the two other flattening functions. The first is a generalization of that obtained by Bishop {\it et al.} \cite{Bishop91}, the second a variant on that discussed above, but emphasizing the small solution rather than the large one.

\section{Cylindrical model examples}
\label{sec:CylinderExamples}

Cylindrical calculations have been carried out to demonstrate the use of this new flattening function. The numerical method used here integrates the Newcomb equation
\begin{equation}
\frac{d}{dr}r\frac{d\psi_m}{dr}-\frac{m^2}{r}\psi_m - \frac{mq\sigma'}{m-nq}\psi_m -\frac{2p'}{B^2_0}\frac{m^2(1-q^2)}{(m-nq)^2}\psi_m =0
\end{equation}   
where $\sigma =J_{||}/B$ is the longitudinal current and, a prime denotes a radial derivative, from the magnetic axis out to the rational surface and then from the plasma boundary into the rational surface. The amplitudes of these two solutions are matched at the rational surface to produce the solution which will have a discontinuity in the first derivative at the rational surface. In this equation we have included toroidal effects i.e. poloidal mode coupling and toroidal curvature which means that $p' \to p'(1-q^2)$, \cite{Bishop91}. This follows from using the large aspect ratio tokamak result for $D_M$, 
\begin{equation}
D_M=\frac{2 p' (1-q^2)}{rB_0^2q'^2}.
\end{equation}
in (\ref{eq:q2final}).

A computer code employing a basis function approach, similar to that used in Connor {\it et al.} \cite{Connor88}, has also been used and produces results consistent with this approach. The case with no flattening function was considered first. The equilibrium used has a safety factor profile given by $q(r) = q_0(1+\lambda r^2/r_a^2)$, where $q_0=1.4$ and $\lambda=1$, which means that the edge safety factor is $q_a=2.8$. The pressure profile is given by $ p(r)=p_0(1-r^2/r_a^2)$. The plasma pressure divided by magnetic pressure, $\beta=0.02$. 

\begin{figure}
\centering
\psfrag{A}{$\log \delta_{rs}$}
\psfrag{B}{$\Delta'_{2,\infty}$}
\includegraphics[width=0.8\textwidth]{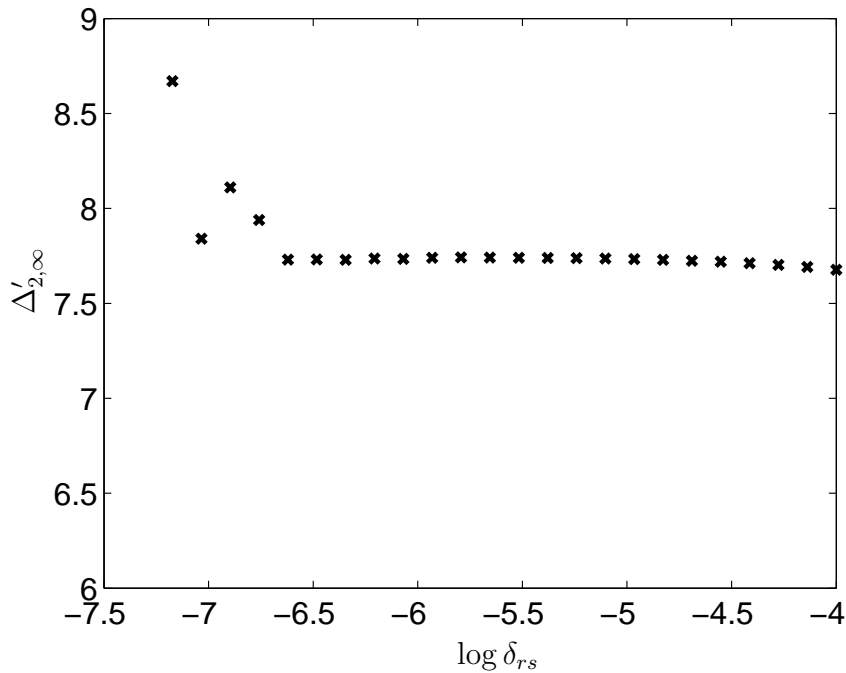}
\caption{$\Delta'_{2,\infty}$ scanned against how close the integration is taken to the rational surface $\delta_{rs}$ on a logarithimic scale. No pressure flattening applied.}
\label{fig:NoFlatPap}
\end{figure}

We consider an $n=1$, $m=2$ mode; figure~\ref{fig:NoFlatPap} shows the $\Delta'_{2,\infty}$ scanned against $\delta_{rs}$, which measures how close the integration is taken to the rational surface. It can be seen that for very small values of $\delta_{rs}$ the result is not well converged.

\begin{figure}
\centering
\psfrag{A}{$\log \hat{\delta}$}
\psfrag{B}{$\Delta'_{2,\infty}$}
\includegraphics[width=0.8\textwidth]{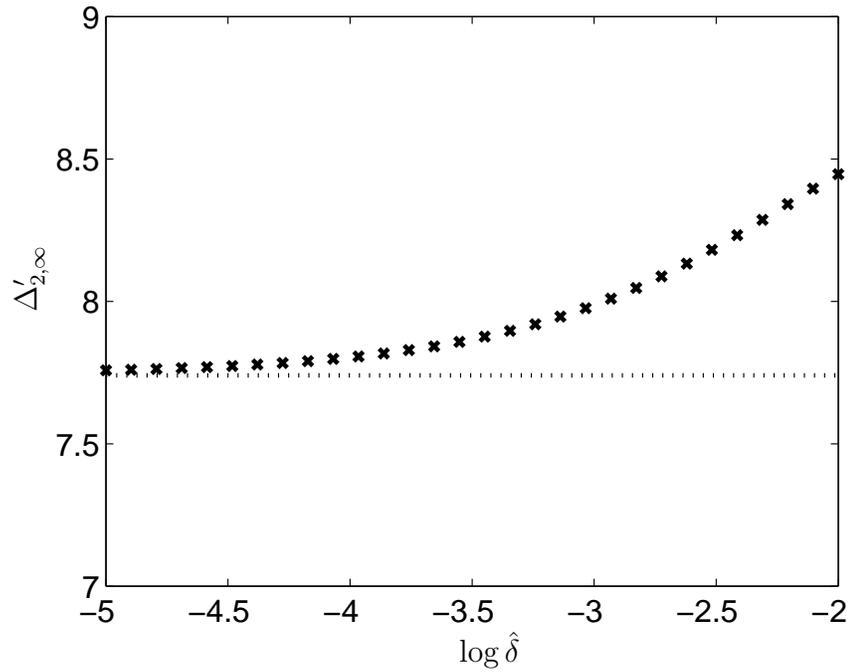}
\caption{$\Delta'_{2,\infty}$ scanned against the width of the flattening function $\hat{\delta}$ on a logarithmic scale. Integration is taken up to $\delta_{rs}=10^{-8}$. The dashed horizontal line is the value calculated without the use of pressure flattening.}
\label{fig:SCCScanft}
\end{figure}

\begin{figure}
\centering
\psfrag{A}{$\log \delta_{rs}$}
\psfrag{B}{$\Delta'_{2,\infty}$}
\includegraphics[width=0.8\textwidth]{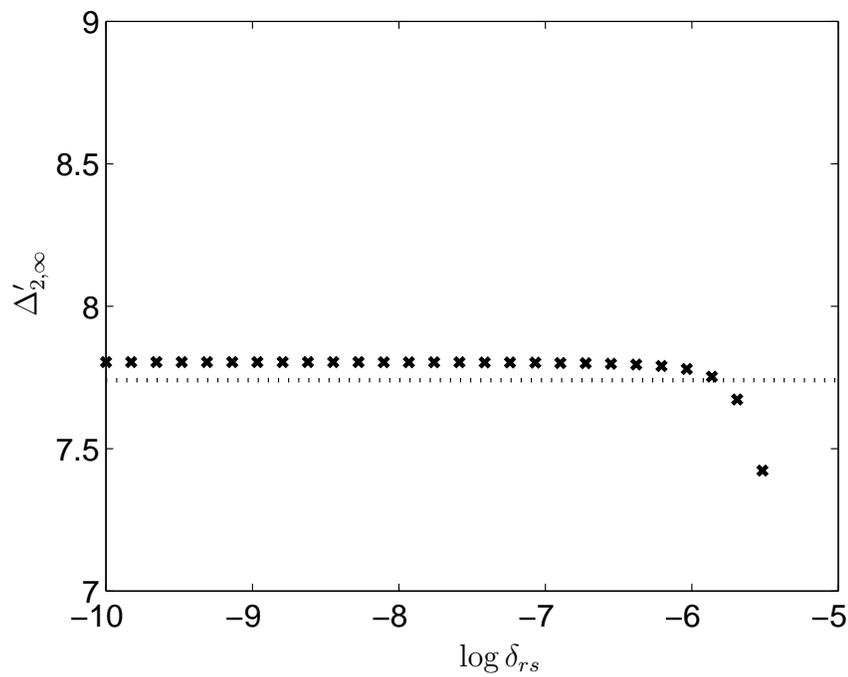}
\caption{$\Delta'_{2,\infty}$ scanned against how close the integration is taken to the rational surface $\delta_{rs}$. Pressure flattening width is fixed at $\hat{\delta}=10^{-4}$. The dashed horizontal line is the value calculated without the use of pressure flattening.}
\label{fig:SCCScanrs}
\end{figure}

Figure~\ref{fig:SCCScanft} shows that $\Delta'_{2,\infty}$  converges to the unflattened result as the flattening width $\hat{\delta}$ decreases. Figure~\ref{fig:SCCScanrs} show that $\Delta'_{2,\infty}$ is well converged for a fixed flattening width as the distance to the rational surface used in the integration, $\delta_{rs}$, is varied. The calculation with pressure flattening displays much better convergence with $\delta_{rs}$ than the case with no pressure flattening because of the absence of the singularity around the rational surface caused by the pressure gradient.  

\section{Toroidal $\Delta'$ calculation}
\label{sec:ToroidalExample}

The motivation for investigating pressure flattening is its potential use for calculating the $\Delta'$ with resistive codes in toroidal geometry. The previous section demonstrated that the method works in cylindrical geometry. In this section we use the T7 code to show that the $\Delta'$ with unflattened pressure can also be recovered from the $\Delta'$ with pressure flattening in toroidal geometry where there is coupling between the poloidal harmonics.   

The equilibrium used for these calculations is similar to that used for the cylindrical examples. We choose a parabolic pressure profile $p(r)=p_0(1-r^2/r_a^2)$ and a safety factor profile $q(r)=q_0(1+\lambda r^2/r_a^2)$ where $q_0=1.4$ and $\lambda =1$ so that the edge safety factor is again $q_a=2.8$. The inverse aspect ratio for these calculations is taken to be $r_a/R_0 =0.1$. 

In a cylinder there is only one second order differential equation to solve for $\psi_m$ and the pressure gradient only appears in this equation once. In a toroidal $\Delta'$ calculation one needs to solve a second order differential equation for each poloidal harmonic included in the calculation. Since the poloidal harmonics are coupled, there is now a set of coupled differential equations to be solved with $p'$ appearing at various points in this set. We will consider two approaches to flattening the pressure in T7. The first is to create a new equilibrium, neighbouring the unflattened equilibrium, which has a flattened pressure. This new pressure profile is then used throughout the coupled equations for the poloidal harmonics. The alternative approach is to flatten the pressure only in the equation for the resonant harmonic, leaving the equations for the other harmonics unchanged. The neighbouring equilibrium approach can be expected to introduce additional changes to $\Delta'$ than those intended: (i) a change in the Shafranov Shift; and (ii) changes to the coupling of harmonics. These effects are examined in Appendix E and are shown to be small.
    
\begin{figure}
\centering
\psfrag{A}{$\log \hat{\delta}$}
\psfrag{B}{$\Delta'_{2,\infty}$}
\includegraphics[width=0.8\textwidth]{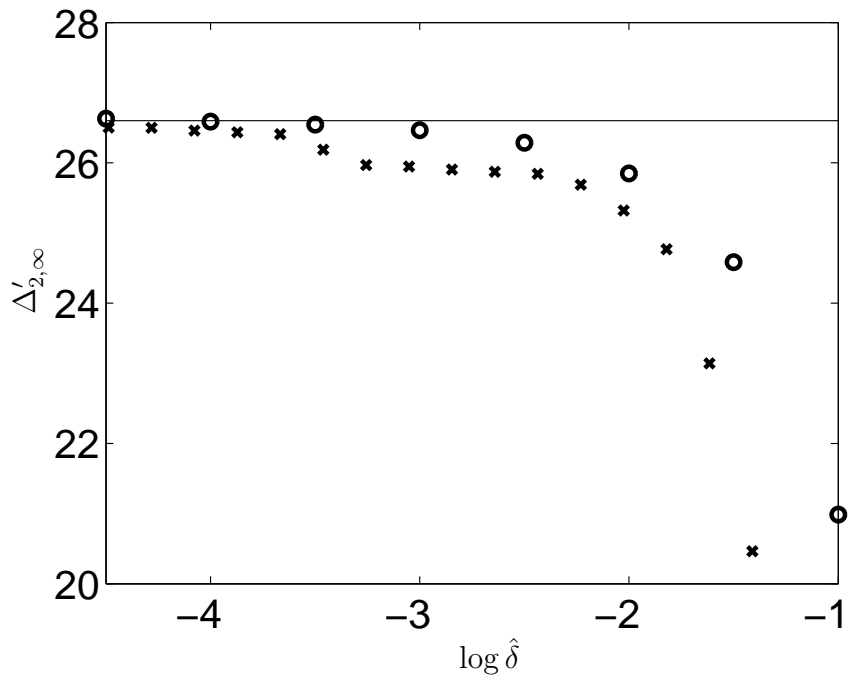}
\caption{$\Delta'_{2,\infty}$ scanned against the width of the flattening function $\hat{\delta}$ on a logarithimic scale. The `$\times$' is the result from the neighbouring equilibrium approach and `o' the results from flattening the pressure only for the resonant harmonic. The horizontal line is the value calculated without the use of pressure flattening.}
\label{fig:T7beta2}
\end{figure}

Figure \ref{fig:T7beta2} compares the results of using T7 to calculate $\Delta'$ with the two approaches. The equilibrium used is as stated above. The `$\times$' show the results from flattening the whole equilibrium and the `o' are the results from just flattening the pressure in the resonant harmonic equation. This plot indicates the latter produces results that converge to the unflattened result (the horizontal line) faster than that when perturbing the whole equilibrium.

\section{Discussion and Conclusions}
\label{sec:Discussion}

In Ref~\cite{Ham12b} we presented a method for calculating the $\Delta'$ in fully toroidal geometry by exploiting resistive MHD codes such as FAR or MARS-F. However, finite pressure gradient and favourable average curvature together lead to the Glasser effect which precludes the straightforward use of these methods to calculate the $\Delta'$; some insights into the reasons for this are given in Appendix A. We have developed a technique here based on flattening the pressure profile in the vicinity of the resonant surface. In Section 2 we derived the equation describing the behaviour of the perturbed flux $\psi_m$ in this region when the pressure profile and the $q$ profile are perturbed about their equilibrium values. This derivation is valid for an arbitrary axisymmetric equilibrium and some details are presented in Appendix B. A local solution of this equation allows one to relate, by an analytic formula, the value of $\Delta'_0$, the value of $\Delta'$ in the presence of flattening, to $\Delta_{\infty}$, its value in the unperturbed equilibrium. When the pressure gradient vanishes at the resonant surface the methods for calculating $\Delta'$ discussed in \cite{Ham12b} are available, so combining these with the analytic relationship above, allows one to calculate $\Delta'$ in the generic case when a pressure gradient is present, overcoming the problems posed by the Glasser effect. (Although the focus of the present work is on the presence of pressure gradients our analysis allows for a perturbation to the $q$ profile and a calculation of the impact of localized current drive on $\Delta'$ is presented in Appendix C.) The general form of the relationship between $\Delta'_0$ and $\Delta'_{\infty}$ was found to be 
\begin{equation}
\Delta'_0=u \hat{\delta}^{-2\nu_L}\Delta'_{\infty} +v/\hat{\delta},
\end{equation}
where $u$ and $v$ are constants. Such an analytic calculation linking $\Delta'_{\infty}$ to $\Delta'_0$ was investigated by Bishop {\it et al.} \cite{Bishop91} but produced a large offset which was a source of numerical difficulty. In Section 3 we designed a form of the pressure perturbation to preserve the amplitude of the large solution throughout the perturbed region. This leads to a link between $\Delta'_0$ and $\Delta'_{\infty}$ that does not involve this large offset and is therefore numerically robust, i.e. $v=0$ for this flattening function. 

We have investigated other methods of flattening the pressure for completeness: the generalization of the method used in Bishop {\it et al.} \cite{Bishop91} and a further method called $D^{(2,l)}$ which we have described in Appendix D. However, both of these methods result in an offset in the relation between the flattened and unflattened $\Delta'$ which is numerically less accurate; it is for this reason we have focused on the flattening function described in Section 3. However, we have included details of $D^{(2,l)}$ because it may be useful in situations where $\Delta'$ is large.

This approach to calculating $\Delta'$ in the presence of pressure was demonstrated in cylindrical geometry in Section 4 and it produced results that converged to the unflattened results. In Section 5 the method was further tested in toroidal geometry using the T7 code. Two approaches to pressure flattening are possible in the torus: perturbing to a neighbouring equilibrium and just perturbing the pressure profile in the equation for the resonant harmonic. Both of these methods converged to the unflattened result as the flattening width was reduced although the approach of just perturbing the pressure for the resonant harmonic worked better in the case investigated. This is because the perturbed equilibrium approach causes changes to the equations for all the harmonics. Although these changes are generally small they result in less favourable convergence than the method that just perturbs the resonant harmonic. We would suggest that this is the method that be implemented in future. Different values of $l$, controlling the sharpness of the flattening functions, were tried; however there was little change in the results for $l\geq 2$. 

It is well known that as the pressure increases the Mercier indices will separate. In T7 these indices are only allowed to separate a small amount before the code gives a warning that the approximation used breaks down. The use of pressure flattening at the rational surface means that this limitation can be circumvented. 

This pressure flattening function approach provides several useful ways to evaluate the true value of $\Delta'$ in a fully toroidal finite beta equilibrium.  If $\Delta'$ is positive, an initial value resistive MHD code, such as FAR, could be used to determine a growth rate for the pressure flattened equilibrium. This growth can be used with a dispersion relation to calculate $\Delta'$. Alternatively, a basis function method using MARS-F as outlined in Ham et al \cite{Ham12}, could be used to determine either $\Delta'_0$, or, $\Delta'_\infty$ directly. 
      
\ack

This work was funded by the RCUK Energy Programme [grant number EP/I501045] and the European Communities under the contract of Association between EURATOM and CCFE.  To obtain further information on the data and models underlying this paper please contact PublicationsManager@ccfe.ac.uk. The views and opinions expressed herein do not necessarily reflect those of the European Commission.

\appendix

\section{The `Glasser effect', shielding and the stationary state}

Glasser, Greene and Johnson \cite{Glasser75} have given the following equations to describe the resistive layer in a toroidal plasma with pressure:

\begin{eqnarray}\label{eq:AppAGGJ}
&&\Psi_{XX} - H \Upsilon_{X}=Q(\Psi-X\Xi)\\
&&Q^2\Xi_{XX}-QX^2\Xi+(E+F)\Upsilon+QX\Psi+H\Psi_X=0 \nonumber\\
&&Q\Upsilon_{XX}-X^2\Upsilon+X\Psi=Q^2\left[(G+FK)\Upsilon -(G-KE)\Xi+KH\Psi_x\right] \nonumber
\end{eqnarray} 
Here $E,F,G,H$ and $K$ are certain flux-surface averaged quantities, $X=(r-r_m)/L_R$ and $Q=q/Q_R$, where $L_R \propto \eta^{1/3}$ is a characteristic resistive length-scale, $q$ is a growth rate with $Q_R \propto \eta^{1/3}$ a resistive interchange growth rate, and $r_m$ is the radius of the resonant flux surface. ($L_R$ corresponds to the width $w$ introduced in Section 2.)

In the construction of basis functions for calculating the tearing mode stability index, $\Delta'$, we propose to take the $Q\to 0$ limit of these equations, corresponding to determing the response to an imposed stationary perturbation at the plasma edge, rather than the growth of a natural tearing mode. We obtain this limit by scaling as follows:
\begin{equation}
QX^2\sim y^2, \quad X\sim \Upsilon \sim Q^{1/2}\Psi, 
\end{equation}
resulting in the equation
\begin{equation}
\frac{d^2}{dy^2}\Psi+\left( D_M +\frac{1}{4}\right)\Psi=0,
\end{equation}
where $D_M=E+F+H$ with the Mercier stability criterion corresponding to $1/4+D_M>0$. This is an `ideal MHD' equation and corresponds to a shielding current layer near the resonant surface of width $X\sim Q^{-1/2}$ as $Q\to 0$.

It is of interest to estimate the resistive decay of such a current by comparing $\partial j/\partial t$ with $\eta\partial^2j/\partial x^2$. We find
\begin{equation}
\eta \frac{\partial^2j}{\partial x^2}/\frac{\partial j}{\partial t} \sim \frac{\eta}{qx^2}\sim \frac{\eta}{QX^2Q_RL^2_R}\sim O(1), 
\end{equation}
i.e. independent of $Q$ and $\eta$, so that this current does not decay. The current occupies a region $y\sim O(1)$, i.e. $x\sim (\eta/q)^{1/2}$ as $q\to 0$.

The origin of Glasser stabilization \cite{Glasser75} is implicit in the cylindrical calculation of Coppi, Greene and Johnson \cite{Coppi66}. They provide a set of equations similar to (\ref{eq:AppAGGJ}) which clearly show that $\Psi \to 0$, i.e. $\delta b\to 0$, near $X=0$ as $Q\to0$. This arises from the parallel force balance equation where the perturbed parallel pressure gradient force arising from $\delta b\cdot \nabla p = \delta b_r(\textup{d}p_0/\textup{d}r)$ in the presence of a finite radial pressure gradient cannot be balanced when the parallel inertia vanishes as $q$ (i.e. $Q$) $\to 0$, since $\nabla_{||} \delta p \propto x \delta p \to 0$.

In fact this result is only true if the ratio of specific heats, $\gamma$, is finite. Otherwise the equation that requires $\Psi\to 0$, i.e. $b_r\to0$, near $X=0$ as $Q\to 0$ plays no role. Actually, if one takes the limit $\gamma \to 0$ in this equation it implies $\Upsilon \propto \Xi$ instead. More physically, ignoring plasma compressibility in the equation of state leads to $\delta b_r$ satisfying the ideal condition, $\delta b_r \propto k_{||} \xi \propto x\xi \to 0$ near $x=0$.

\section{Equilibrium relations for derivation of tearing mode equation with pressure perturbation}

In this section we follow the work of Greene and Chance \cite{Greene81} in deriving the essential constraint among the various perturbed profiles when an equilibrium is locally modified. Following Connor {\it et al.} \cite{Connor88} we represent the magnetic field in the form
\begin{equation}
\label{eq:Bapp}
{\bf B} = R_0B_0(f(r) \nabla \phi \times \nabla r + g(r) \nabla \phi)
\end{equation}
and employ field line straightened coordinates $r, \theta, \phi$. In (\ref{eq:Bapp}) $r$ is a magnetic surface coordinate, $\phi$ the toroidal angle, $R_0$ is the major radius at the magnetic axis, and $B_0$ the vacuum magnetic field there. The poloidal angle $\theta$ is chosen so that field line trajectories are straight, therefore $\frac{d\phi}{d\theta}=q(r)$, in these coordinates. The jacobian is then given by
\begin{equation}
j=(\nabla r \times \nabla \theta \cdot \nabla \phi)^{-1} =\frac{rR^2}{R_0}
\end{equation} 
With this choice of coordinates the safety factor can be expressed in terms of the magnetic field variables as
\begin{equation}
\label{eq:qdef}
q(r) =\frac{rg}{R_0f}
\end{equation}
and the Grad-Shafranov equation takes the form 
\begin{equation}
\frac{1}{r}\frac{\partial}{\partial r}( rf|\nabla r|^2) +f\frac{\partial}{\partial \theta}(\nabla \theta \cdot \nabla r) +\frac{1}{f} \left( gg' +\frac{R^2}{R^2_0 B^2_0}p' \right) =0,
\end{equation}
where $'$ means the derivative with respect to $r$.
We consider perturbations about an equilibrium described by
\begin{equation}
p=p^{(0)}(r), \quad f=f^{(0)}(r), \quad g=g^{(0)}(r)
\end{equation}
such that
\begin{eqnarray}
\label{eq:pfg}
p&=& p^{(0)}(r) +\hat{\delta} p^{(1)}(t)\nonumber\\
f&=& f^{(0)}(r) +\hat{\delta} f^{(1)}(t)\nonumber\\
g&=& g^{(0)}(r) +\hat{\delta} g^{(1)}(t)
\end{eqnarray}
where $t=(r-r_m)/\delta$, with $\delta \ll r_m$ providing a measure of the localization about $r_m$. To calculate the perturbations to the metric of the flux surface coordinates $(r, \theta, \phi)$ induced by the profile perturbations (\ref{eq:pfg}), we represent the $(R,Z) \to (r,\theta)$ transformation in the form:-
\begin{eqnarray}
\label{eq:Rapp}
R&=& R^{(0)}(r,\theta)=\hat{\delta} R^{(1)}(t,\theta)\\
\label{eq:Zapp}Z&=& Z^{(0)}(r,\theta)=\hat{\delta} Z^{(1)}(t,\theta)
\end{eqnarray} 
The perturbations in the metric can then be determined from the relations
\begin{eqnarray}
\label{eq:gradr2}
|\nabla r|^2 &=&\left( \frac{\partial r}{\partial R}\right)^2 +\left(\frac{\partial r}{\partial Z}\right)^2 = \frac{R^2_0}{r^2R^2}\left[ \left(\frac{\partial R}{\partial \theta}\right)^2+ \left(\frac{\partial Z}{\partial \theta}\right)^2 \right]\\
\label{eq:gradtr}\nabla \theta \cdot \nabla r &=& -\frac{R^2_0}{r^2R^2}\left[\frac{\partial R}{\partial \theta}\frac{\partial R}{\partial r} +\frac{\partial Z}{\partial \theta}\frac{\partial Z}{\partial r}  \right]
\end{eqnarray}
by substituting (\ref{eq:Rapp}) and (\ref{eq:Zapp}) into (\ref{eq:gradr2}) and (\ref{eq:gradtr}) and using $\frac{\partial}{\partial r}\to \frac{\partial}{\partial r}+\frac{1}{r_m\delta}\frac{\partial}{\partial t}$. First, using the result for the jacobian of $(R,Z)\to (r,\theta)$, 

\begin{equation}
\frac{r R}{R_0}= \left( \frac{\partial R}{\partial \theta}\frac{\partial Z}{\partial r} -\frac{\partial R}{\partial r}\frac{\partial Z}{\partial \theta}  \right)
\end{equation} 
we obtain
\begin{equation}
\frac{\partial Z^{(1)}}{\partial t}\frac{\partial R^{(0)}}{\partial \theta} -\frac{\partial R^{(1)}}{\partial t}\frac{\partial Z^{(0)}}{\partial \theta}=0
\end{equation}
which constrains $Z^{(1)}$, $R^{(1)}$ to the forms
\begin{equation}
R^{(1)}=h(t,\theta)\frac{\partial R^{(0)}}{\partial \theta}; \quad Z^{(1)}=h(t,\theta)\frac{\partial Z^{(0)}}{\partial \theta}.
\end{equation}
Equations (\ref{eq:gradr2}) and (\ref{eq:gradtr}) then yield the results 
\begin{eqnarray}
|\nabla r|^2&=& |\nabla r|^{(0)2} \left \{ 1+ 2\hat{\delta} \frac{\partial h}{\partial \theta} + \frac{\hat{\delta} h}{|\nabla r|^{(0)2}} \frac{\partial }{\partial \theta} |\nabla r|^{(0)2} \right \} \\
\nabla r \cdot \nabla \theta &=& (\nabla r \cdot \nabla \theta )^{(0)} - \frac{r_m}{r}|\nabla r|^{(0)2}\frac{\partial h}{\partial t} +O(\hat{\delta}) 
\end{eqnarray} 
Inserting these results in the perturbed Grad-Shafranov equation we obtain, to zero order in $\hat{\delta}$, 
\begin{equation}
\label{eq:pGS}
\frac{f^{(0)}}{r}|\nabla r|^{(0)2}\frac{df^{(1)}}{dt}+ rf^{(0)2}|\nabla r|^{(0)2} \frac{\partial^2 h}{\partial r \partial \theta} 
+ \frac{g^{(0)}}{r}\frac{d g^{(1)}}{ dt} + \frac{R^2}{R^2_0 B^2_0r}\frac{d p^{(1)}}{ dt}=0
\end{equation}
The solubility condition for (\ref{eq:pGS}) is obtained by annihilating the term involving $h(t,\theta)$. Thus writing $\langle A \rangle \equiv \frac{1}{2 \pi}\oint A d \theta$, we obtain
\begin{equation}
\label{eq:fAppB}
f^{(0)}\frac{df^{(1)}}{dt}+g^{(0)}\frac{dg^{(1)}}{dt}\left \langle \frac{1}{|\nabla r|^{(0)2}} \right \rangle + \frac{1}{B^2_0}\frac{dp^{(1)}}{dt} \left \langle \frac{R^2}{R^2_0 |\nabla r|^{(0)2}} \right \rangle =0
\end{equation}
and finally the solution for $h(r,\theta)$ in terms of $f^{(1)}$, $g^{(1)}$, $p^{(1)}$, and the unperturbed equilibrium quantities.

Since it is convenient for the tearing mode analysis to describe the equilibrium in terms of pressure and safety factor profiles, we eliminate $f^{(1)\prime}$, $g^{(1)\prime}$, by introducing $q^{(1)}$, with,
\begin{equation}
\label{eq:qAppB}
q=q^{(0)}(r) +\hat{\delta} q^{(1)}(t)
\end{equation}
so that, using (\ref{eq:qdef}) we obtain
\begin{equation}
\label{eq:oqAppB}
\frac{1}{q^{(0)}}\frac{d q^{(1)}}{dt}= \frac{1}{g^{(0)}}\frac{d g^{(1)}}{dt} - \frac{1}{f^{(0)}}\frac{d f^{(1)}}{dt}
\end{equation}
Thus using (\ref{eq:fAppB}) and (\ref{eq:oqAppB}) in (\ref{eq:pGS}) we finally obtain
\begin{eqnarray}
\label{eq:AppBFinal}
&&\left \langle \frac{B^2}{B^2_0} \frac{R^2}{R^2_0}\frac{1}{|\nabla r|^2} \right \rangle \frac{\partial^2 h}{\partial t \partial \theta}=
\frac{g^2}{q}\frac{d q^{(1)}}{dt} \left [\left \langle \frac{1}{|\nabla r|^2}\right \rangle - \frac{1}{|\nabla r|^2}  \right]\\
&& + \frac{1}{B_0^2 f^2}\frac{d p^{(1)}}{dt} \left \{ \frac{B^2}{B^2_0} \frac{R^2}{R^2_0}\frac{1}{|\nabla r|^2} \left \langle \frac{R^2}{R^2_0 |\nabla r|^2} \right \rangle - \frac{R^2}{R^2_0|\nabla r|^2} \left \langle \frac{B^2R^2}{R^2_0 B^2_0 |\nabla r|^2} \right \rangle    \right \} \nonumber
\end{eqnarray}
These perturbed equilibrium relations provide the necessary information for the stability analysis of tearing modes in Section 2. 

\section{Stabilization by local current drive}

In this paper we have concentrated on perturbations to the pressure gradient in (\ref{eq:q2final}). However, in this appendix we consider the opposite limit of (\ref{eq:q2final}) in which the pressure is neglected, but local current drive produces a change $q(t)$ of arbitrary form in the safety factor. The equation governing stability now becomes
\begin{equation}
(\Delta q)\frac{d^2\psi_m}{dt^2}=\psi_m\frac{d^2\Delta q}{dt^2}
\end{equation}
and locally has the general solution 
\begin{equation}
\label{eq:AppCpsi}
\psi_m=c\Delta q(t) +\Delta q(t) \int^t \frac{dy'}{(\Delta q)^2}
\end{equation}
where $c$ takes different values $c^-$, and $c^+$, for the solutions to the left and right of the resonant surface.

Expressions can now readily be derived from (\ref{eq:AppCpsi}) for $\Delta'_0$ (by taking the limit $t \to 0$) and $\Delta'_{\infty}$ (by taking the limit $|t| \to \infty$, where $q^{(1)}(t)\to 0$). For convenience we introduce the following notation:-
\begin{eqnarray}
q'_0&=&\lim_{t\to 0}\left(\frac{d}{dr}+ \frac{1}{\delta}\frac{d}{dt}\right)(\hat{\delta} \Delta q)\nonumber \\
q'_{\infty}&=&\lim_{t\to \infty}\left(\frac{d}{dr}+ \frac{1}{\delta}\frac{d}{dt}\right)(\hat{\delta} \Delta q)
\equiv q'^{(0)}(r_m)\nonumber \\
s_0&=&\frac{r_m q'_0}{q}, \quad s_{\infty}=\frac{r_m q'_{\infty}}{q}
\end{eqnarray}
Thus we find
\begin{equation}
\psi^-(t)=\Delta q\left(c^-+ \int^t_{-\infty} \frac{dt'}{(\Delta q)^2}  \right)
\end{equation}
with asymptotic forms at $t \to -\infty$ and $t \to 0$ given by
\begin{eqnarray}
\psi^-_{\infty} &\simeq & q'_{\infty}\left(c^-t -\frac{1}{(q'_{\infty})^2} \right)\\
\psi^-_{0} &\simeq & q'_0t \left (c^- +\int^{-1}_{-\infty}dt \left [ \frac{1}{(\Delta q)^2} -\frac{1}{(q't)^2}\right ] \right) \\ 
&+& \left( \int^{0}_{-1} dt \left [\frac{1}{(\Delta q)^2}-\frac{1}{(q'_0t)^2}+\frac{2q''_0}{(q'_0)^3t}  \right] -\frac{1}{(q'_0)^2 t} -\frac{2q''_0}{(q'_0)^3}\ln |t|  \right)
\end{eqnarray}
and similar expressions for $\psi^+_{\infty}$ and $\psi^+_{0}$. Making use of these expressions to construct $\Delta'_0$ and $\Delta'_{\infty}$, we obtain
\begin{equation}
\label{eq:DeltaAppC}
\Delta'_0 = \frac{s^2_0}{s^2_{\infty}}\Delta'_{\infty}+ \frac{1}{\hat{\delta}}P\int^{\infty}_{-\infty} dt \left [ \frac{s^2_0 q^2}{(\Delta q(t))^2} -\frac{1}{t^2} \right ].
\end{equation}

Equation (\ref{eq:DeltaAppC}) is the main result of this appendix. In the limit of weak current perturbations $s_0 \to s_{\infty} \equiv s$, $\Delta q \to sqx+\frac{1}{2}x^2(\Delta q)''$ with $(\Delta q)'' \ll s_0 q$, (\ref{eq:DeltaAppC}) reduces to the expressions obtained in \cite{Zakharov89, Westerhof90}:
\begin{equation}
\Delta'_0 = \Delta'_{\infty} - P \int^{\infty}_{-\infty} r_m\frac{(\Delta q)''(x)}{sq}\frac{dx}{x}
\end{equation}
with $ -\frac{(\Delta q)''}{q^2}\propto J'_{||}(x)$, where $J_{||}$ is the parallel current.

Returning to (\ref{eq:DeltaAppC}) we consider an application to $m=1$, $n=1$ stabilization. In this case $\Delta'_{\infty}$ is given by
\begin{equation}
\Delta' \equiv \Delta'_{1/1}=\frac{s^2_{\infty}}{\epsilon^2_1\delta W_{T}}
\end{equation}
where $s_{\infty}\equiv r_1q'(r_1)$ is the shear at the $q=1$ surface in the unperturbed equilibrium, $\epsilon_1 = r_1/R_0$ and $\delta W_T$ is the ideal MHD energy integral obtained originally by Bussac {\it et al.} \cite{Bussac75}. As an elementary example we consider perturbations of the form
\begin{equation}
\Delta q(t) = s_0 t \frac{\left(1 + \left(\frac{s_{\infty}}{s_0}\right)^2 t^2 \right )^{1/2}}{(1+t^2)^{1/2}}
\end{equation}
so that (\ref{eq:DeltaAppC}) yields the result.
\begin{equation}
\Delta'_0=\frac{s^2_0}{s^2_{\infty}}\frac{s^2_{\infty}}{\epsilon^2_1 \delta W_T} + \frac{\pi}{\hat{\delta}}\left ( \frac{s_0}{s_{\infty}} -\frac{s_{\infty}}{s_0} \right )
\end{equation}
Thus marginal stability $(\Delta'_0 \to 0)$ is achieved by reducing the shear $s_0$ at the singular layer to the value
\begin{equation}
s_0 \sim \left( \pi \frac{\epsilon^2_1 s_{\infty}}{\hat{\delta}}\delta W_T \right)^{1/3}
\end{equation}
This simple result serves to indicate how local current perturbations, which are strong enough to modify the shear at the $q=1$ surface, can totally stabilize the $m=1$ tearing mode.

\section{Other flattening functions}
\label{sec:3flatfn}

Three pressure flattening functions are discussed in this appendix: $D^{(1,l)}$, $D^{(2,l)}$ and $D^{(B,l)}$. Bishop {\it et al.}~\cite{Bishop91} investigated the function $D^{(B,1)}$ analytically, however we generalize this original treatment to arbitrary $l$.

\subsection{Flattening function $D^{(2,l)}$}

Instead of prescribing a ``nice structure'' for the large solution as described in the main paper, we now do so for the small solution.  At first sight this appears an irrelevance to the construction of a fully reconnected basis function, but it proves otherwise. Following the ideas underlying (\ref{eq:psiL}) we choose one solution to be:
\begin{equation}
\psi(t)=|t|(1+|t|^{2l})^{\frac{\nu_S-1}{2l}}
\end{equation}
where
\begin{equation}
\nu_S=\frac{1}{2}(1+\sqrt{1+4D_\infty})
\end{equation}
is the Mercier index of the small solution. As before, the flattening function $D(t)$ which achieves this can be calculated and is:
\begin{equation}
D^{(2,l)}(t)=\frac{t^{4l}}{(1+t^{2l})^2}\left [ D_\infty + \frac{(2l+1)(\nu_S-1)}{t^{2l}}\right ]
\end{equation}   
and as in Section 3, a second solution to (\ref{eq:Start}) can be generated. This time we choose the arbitrary constant in order to make this additional solution pure large solution at large $t$ values. Specifically it is given by:
\begin{equation}
\psi(t)=|t|(1+|t|^{2l})^{\frac{\nu_S-1}{2l}} (2 \nu_S-1)\int_{|t|}^\infty \frac{dx}{x^2(1+x^{2l})^{\frac{\nu_S-1}{l}}}
\end{equation}
so that as $|t|\rightarrow \infty$, $\psi \sim |t|^{\nu_L}$, i.e. pure large solution.
     
We again consider how an arbitrary mix of large and small solutions behaves as it passes through this choice of flattening function.  i.e. through $D^{(2,l)}(t)$, rather than $D^{(1,l)}(t)$. Assuming, as before, a perturbed flux function  of the form (\ref{eq:psimix}):
\begin{equation}
\psi(t) \sim \alpha \hat{\delta}^{\nu_L} |t|^{\nu_L} + \beta \hat{\delta}^{\nu_S} |t|^{\nu_S}
\end{equation}
at large values of the local variable $|t|$, this takes the form:
\begin{equation}
\label{eq:psiD2}
\psi  \sim \alpha (2 \nu_S-1)\hat{\delta}^{\nu_L}  + \left [ \beta \hat{\delta}^{\nu_S-1} - \alpha (2 \nu_S-1) I_1 \hat{\delta}^{\nu_L-1} \right ] |x|
\end{equation}
for $|t|\ll1$, where:
\begin{equation}
I_1 =\int_0 ^\infty \frac{dt}{t^2} \left [ 1-\frac{1}{(1+|t|^{2l})^{\frac{\nu_S-1}{2l}}}\right ] =\frac{\Gamma(1-\frac{1}{2l} )\Gamma(\frac{\nu_S}{2l})}{\Gamma(\frac{\nu_S -1}{2l})}
\end{equation} 
Thus, in this case the large solution again emerges with increased amplitude, but without contamination generated by the small solution, as it passes through the flattening zone. Consequently, using $D^{(2,l)}(t)$, rather than $D^{(1,l)}(t)$,  as the flattening function may be more accurate when the required $\Delta^\prime$ is large.  The fully reconnected amplitude (i.e. the magnitude of $\alpha$ in (\ref{eq:psiD2}) can then be obtained from a code like MARS-F for example in the form
\begin{equation}
\alpha=\frac{\psi_m(0)}{(1-2 \nu_L)\hat{\delta}^{\nu_L}}
\end{equation}
and we obtain
\begin{equation}
\label{eq:D0AppD}
\Delta'_0=\frac{\Delta'_\infty~\hat{\delta}^{-2\nu_L}}{1-2 \nu_L}-\frac{2I_1}{\hat{\delta}}
\end{equation}
and in this case (\ref{eq:D0AppD}) is exact.
     
\subsection{Generalized Bishop Flattening Function $D^{(B,l)}$} 
 
Analytic solution, in terms of the Hypergeometric function, is also possible for the generalised Bishop flattening function
\begin{equation}
D^{(B,l)}(X)=\frac{t^{2l}}{1+t^{2l}}.
\end{equation}
In this case the solution of the form,
\begin{equation}
\label{eq:nBispsi}
\psi(t)=c_1 + c_2 |t|
\end{equation}
for $|t| \ll 1$ has the following asymptotic structure at $|t|\gg 1$:
\begin{eqnarray}
\label{eq:nBis}
\psi & \rightarrow & |t|^{\nu_L} \left [ c_1 \frac{\Gamma(1-\frac{1}{2l}) \Gamma(\frac{\nu_L-\nu_S}{2l})}{\Gamma(\frac{-\nu_S}{2l})\Gamma(1-\frac{\hat{\delta}}{2l})}+c_2 \frac{\Gamma(1+\frac{1}{2l})\Gamma(\frac{\nu_L-\nu_S}{2l})}{\Gamma(\frac{\nu_L}{2l})\Gamma(1+\frac{\nu_L}{2l})} \right ] \nonumber \\
& + & |t|^{\nu_S} \left [ c_1 \frac{\Gamma(1-\frac{1}{2l}) \Gamma(\frac{\nu_S-\nu_L}{2l})}{\Gamma(\frac{-\nu_L}{2l})\Gamma(1-\frac{\nu_L}{2l})}+c_2 \frac{\Gamma(1+\frac{1}{2l})\Gamma(\frac{\nu_S-\nu_L}{2l})}{\Gamma(\frac{\nu_S}{2l})\Gamma(1+\frac{\nu_S}{2l})} \right ]
\end{eqnarray}
With this information, and using MARS-F, for example, to construct a fully reconnected basis function and to give values for $c_1$ and $c_2$ for such a function, the corresponding amplitude of the large solution, $\propto |t|^{\nu_L}$, at large $t$ can be inferred from (\ref{eq:nBis}).  Equations (\ref{eq:nBis}) and (\ref{eq:nBispsi}) can also be used to derive the relation between $\Delta'_\infty$ and $\Delta'_0$, \cite{Bishop91}. The result is:
\begin{equation}
\label{eq:DBisGen}
\Delta^{\prime}_0 =\hat{\delta}^{-2\nu_L} \Delta'_\infty R - \frac{Q}{\hat{\delta}},
\end{equation}
where
\begin{eqnarray}
\hspace*{-2.0cm}Q(d,l)\hspace*{-0.0cm}&=&\hspace*{-0.0cm} 2 \frac{\Gamma(1-\frac{1}{2l})}{\Gamma(1+\frac{1}{2l})} \frac{\Gamma(\frac{\nu_S}{2l})\Gamma(1+\frac{\nu_S}{2l})}{\Gamma(-\frac{\nu_L}{2l})\Gamma(1-\frac{\nu_L}{2l})} ,\nonumber \\
\hspace*{-2.0cm}R(d,l)\hspace*{-0.0cm}&=&\hspace*{-0.0cm}\left [ \frac{\Gamma(\frac{\nu_L-\nu_S}{2l})}{\Gamma(\frac{\nu_S-\nu_L}{2l})}\right ] \frac{\Gamma(1-\frac{1}{2l})}{\Gamma(1+\frac{1}{2l})} \left [ \frac{\Gamma(\frac{\nu_S}{2l}) \Gamma(1+\frac{\nu_S}{2l})}{\Gamma(\frac{-\nu_S}{2l})\Gamma(1-\frac{\nu_S}{2l})}\right ] \left [ \frac{\Gamma(\frac{\nu_S}{2l})\Gamma(1+\frac{\nu_S}{2l})\Gamma(\frac{-\nu_S}{2l})\Gamma(1-\frac{\nu_S}{2l})}{\Gamma(\frac{\nu_L}{2l})\Gamma(1+\frac{\nu_L}{2l})\Gamma(\frac{-\nu_L}{2l})\Gamma(1-\frac{\nu_L}{2l})} -1 ~\right ],\nonumber
\end{eqnarray}
which reduces to the expression found by Bishop {\it et al.} \cite{Bishop91} in the case $l=1$.

\subsection{Summary}

Finally we summarise the relationship between $\Delta'_{\infty}$ and $\Delta'_0$ for the various flattening functions when $D_{M\infty} \ll 1$
 
\begin{eqnarray}
\Delta^{\prime(1,l)}_{0}&\sim& \hat{\delta}^{-2\nu_L} \Delta^\prime_\infty, \\
\Delta^{\prime(2,l)}_{0}& \sim&\hat{\delta}^{-2\nu_L} \Delta^{\prime}_\infty -\frac{2D_{M\infty}}{\hat{\delta}} \frac{\pi}{2l}\mathrm{cosec}(\frac{\pi}{2l}),\\
\label{eq:AppBis}\Delta^{\prime(B,l)}_{0}&\sim& \hat{\delta}^{-2\nu_L}\Delta^{\prime}_\infty-\frac{\pi D_{M\infty}}{2\hat{\delta}}
\end{eqnarray}
where (\ref{eq:AppBis}) refers to the Bishop {\it et al.} \cite{Bishop91} choice of flattening function.

\section{Effects of Neighbouring Equilibrium Approach} 

\subsection{Drop in core pressure}

Any choice of flattening function, $D(t)$, which eliminates the pressure gradient across a small,  $O(\delta)$, zone of the equilibrium will result in a small reduction of the central $\beta$, $\beta_0$, value of the equilibrium, with a consequent slight reduction in pressure driven coupling effects (due to a reduced Shafranov shift) in the ideal outer region. The magnitude of this effect can be calculated by evaluating the integral:
\begin{equation}
I_{\beta}^{(j)}= \int_{-\infty}  ^{+ \infty}(1-\frac{D^{(j)}(t)}{D_{M\infty}}) dt, \qquad j=1,2
\end{equation}
so that,
\begin{equation}
\frac{\delta \beta_0}{\beta_0}=\frac{\delta}{r_m}  I_{\beta}.
\end{equation}
The integral $I_{\beta}^{(j)}$ has been evaluated in general, with the result:
\begin{equation}
I_{\beta}^{(j)}=\frac{\pi}{2l^2} \mathrm{cosec}\left(\frac{\pi}{2l}\right )\left [ (2l+1)+\frac{2l-1}{1-\nu_L}\right ]
\end{equation}
Thus $\delta \beta_0/\beta_0$ is small of $O(\delta/r_m)$.

\subsection{$l\to \infty$ limit of the flattening functions}

As $l \rightarrow \infty$, $D_1$ takes the form;
\begin{equation}
D^{(1,l)}(t) = D^{(B,l)}(t)+\nu_L \lbrack  \delta(t-1)+\delta(t+1)\rbrack
\end{equation}
where, as before, $\nu_L$ and $\nu_S$ are the Mercier indices of the large amd small solutions respectively and $\delta(x)$ is the Dirac $\delta$ function. In toroidal configurations the $\delta$ functions represent current sheets arising from the Pfirch-Schl\"uter current and will drive harmonic coupling (large jumps in the sideband harmonics, $\psi_{m \pm1}$ ) which will be important unless their effects at $t=+1$ and $t=-1$ cancel.  

We estimate this toroidal coupling, i.e. the jumps in the sideband harmonics $\psi_{m \pm1}$,  caused by the $\delta$ functions at $t=\pm1$ in $D^{(1,l)}(t)$. This is determined by, e.g. (A.19) of Fitzpatrick {\it et al.} \cite{Fitzpatrick93}, namely;
\begin{equation}
r\frac{d \psi_{m+1}}{dr}=\frac{M_{m+1}^{m} \psi_m}{m-nq}+...,
\end{equation}
with,
\begin{equation}
M_{m+1}^m=(m+1)(m+1-nq)\frac{R p^{\prime}q^2}{B_0^2}+...
\end{equation}
Now, integrating (E.5) through the $\delta$ functions in $p'$ at $t=1$ and at $t=-1$, for the case of $D(t)$, one finds:
\begin{eqnarray}
\lbrack \psi_{m+1}\rbrack_{-1-}^{-1+}&=&  \frac{\delta \nu_L}{2 D_{M\infty}}(m+1)\frac{(1+nq^\prime \delta)(\alpha+\alpha^\prime \delta)}{nq^\prime \delta+nq^{\prime \prime}\delta^2/2} \int_{-1-}^{-1+}\delta(t+1)dt,\\
\lbrack \psi_{m-1}\rbrack_{1-}^{1+}  &=& -\frac{\delta \nu_L}{2D_{M\infty}}(m+1)\frac{(1-nq^\prime \delta)(\alpha-\alpha^\prime \delta)}{nq^\prime \delta-nq^{\prime \prime}\delta^2/2} \int_{1-}^{1+}\delta(t-1)dt,
\end{eqnarray}
Consequently the combined effect of both $ \delta$ functions is a small jump in the $(m\pm1)$  sideband harmonics  given by:
\begin{equation}
\lbrack \psi_{m\pm 1} \rbrack_{r_s-}^{r_s +} \sim \frac{\nu_L}{D_{M\infty}}\frac{ \alpha}{s}\frac{ \delta}{ r_s}.
\end{equation}

\section*{References}

\bibliography{T7bib}{}

\begin{thebibliography}{10}

\bibitem{LaHaye06}
La Haye R J 2006 {\em Phys. Plasmas}  {\bf 13} 055501

\bibitem{Furth63}
Furth H P, Killeen J, and Rosenbluth M N 1963 {\em Phys. Fluids} {\bf 6} 459 

\bibitem{Connor88}
Connor J W {\it et al} 1988 {\em Phys. Fluids} {\bf 31} 577

\bibitem{Fitzpatrick93}
Fitzpatrick R {\it et al} 1993 {\em Nucl. Fusion}, {\bf 33} 1533

\bibitem{Cowley88}
Cowley S C and Hastie R J 1988 {\em Phys. Fluids}, {\bf 31} 426

\bibitem{Pletzer91}
Pletzer A and Dewar R L 1991 {\em J. Plasma Phys.} {\bf 45} 427

\bibitem{Brennan02}
Brennan D P {\it et~al} 2002 {\em Phys. Plasmas} {\bf 9} 2998

\bibitem{Charlton86}
Charlton L A {\it et al} 1986 {\em J. Comput. Phys.} {\bf 63} 107

\bibitem{Hender87}
Hender T C, Hastie R J, and Robinson D C 1987 {\em Nucl. Fusion} {\bf 27} 1389

\bibitem{Liu00}
Liu Y Q {\it et al} 2000 {\em Phys. Plasmas} {\bf 7} 3681

\bibitem{Ham12b}
Ham C J {\it et al} 2012 {\em Plasma Phys. Control. Fusion} {\bf 54} 105014

\bibitem{Newcomb60}
Newcomb W A 1960 {\em Ann. Phys.}  {\bf 10} 232

\bibitem{Glasser75}
Glasser A H, Greene J M, and Johnson J L 1975 {\em Phys. Fluids} {\bf 18} 875

\bibitem{Liu12}
Liu Y Q {\it et al} 2012 {\em Phys. Plasmas} {\bf 19} 072509

\bibitem{Coppi66}
Coppi B, Greene J M, and Johnson J L 1966 {\em Nucl. Fusion} {\bf 6} 101

\bibitem{Bishop91}
Bishop C M, Connor J W, Hastie R J, and Cowley S C 1991 {\em Plasma Phys. Control. Fusion} {\bf 33} 389

\bibitem{Mercier60}
Mercier C 1960 {\em Nucl. Fusion} {\bf 1} 47

\bibitem{Ham12}
Ham C J {\it et al} 2012 {\em Plasma Phys. Control. Fusion}  {\bf 54} 025009

\bibitem{Greene81}
Greene J M and Chance M S 1981 {\em Nucl. Fusion} {\bf 21} 483

\bibitem{Zakharov89}
Zakharov L E and Subbotin A A 1989  ITER-IL-PH-11-9-52

\bibitem{Westerhof90}
Westerhof E 1990 {\em Nucl. Fusion} {\bf 30} 1143

\bibitem{Bussac75}
Bussac M N {\it et~al} 1975 {\em Phys. Rev. Lett}  {\bf 35} 1638

\end{thebibliography}
\bibliographystyle{unsrt}
\end{document}